\documentclass[%
 reprint,
 amsmath,amssymb,
 aps,
]{revtex4-2}

\usepackage{graphicx}
\usepackage{dcolumn}
\usepackage{bm}
\usepackage{xurl}
\usepackage{placeins}

\begin{document}

\preprint{APS/123-QED}

\title{Modeling Quantum Billiards with the Finite Element Method:\\Searching for Quantum Scarring Candidates}

\author{Daniel J. Pierce}
\author{Renuka Rajapakse}%
\affiliation{%
 Department of Physics \\
University of Massachusetts Dartmouth \\
Dartmouth, MA
}%


\date{\today}

\begin{abstract}
An electron in quantum confinement takes on a discrete energy spectrum which is defined based on the solution to the Schr{\"o}dinger Equation for a given potential. Well defined closed-form energy spectra are known for the particle in a box, circular potential, quarter circle potential, and an equilateral triangle. A closed-form solution for more complex shapes may not be known, but numerical methods can be used to find an approximate solution. In this research, an application of the Finite Element Method (FEM) in Wolfram Mathematica is presented and applied to Quantum Billiards with a variety of geometries. To assess the accuracy of the method, the computed energy states are analyzed in the limit of a polygon with an increasing number of sides, the numerical results are validated against analytical solutions for geometries with known exact forms, and a standard convergence test is conducted. The FEM results closely match analytical solutions for known potentials, demonstrating its high accuracy. For high energy index $n$, quantum scarring may emerge for certain geometries. The nature of quantum scarring and its presence in the computed models is also investigated qualitatively. 
\end{abstract}

\maketitle


\section{\label{sec:level1} Introduction}

Many-body quantum systems are a requirement for the implementation of practical, large-scale, quantum computers in the modern age. Controlling such systems is important to minimize decoherence \cite{Gross2017}. It has recently been shown that strong disorder causes suppression of entanglement and allows a memory of the system to persist for a long time \cite{Abanin2019}. Identifying states where dynamics can be controlled externally is an important area of research. When considering bound states of a classically chaotic Hamiltionian system, certain periodic orbits create ``scars", or areas of higher density corresponding to classical orbits \cite{Heller-84}. Quantum billiard systems display periodic orbits that for some eigenstates very closely resemble classical orbits.

Quantum Computers can be designed with quantum dots, made from semiconductor materials, as quantum bits \cite{Harvey-22}. Quantum dots in themselves are not many-body quantum systems, but collectively their implementation as many qubits \textit{is} a many-body quantum system. Quantum billiards are idealized models of quantum dots and provide a powerful framework for understanding the quantum dynamics of electron confinement \cite{Porter-01}. The study of quantum billiard systems offers valuable insight into wave function structure, quantum interference, and the emergence of phenomena that continue to inform our understanding of more complex many-body quantum systems \cite{Ravnik-21}.

The primary focus of this work was to accurately model quantum billiard systems with the Finite Element Method (FEM). In this paper, the eigenvalues and eigenfunctions found for each geometric region are presented, and their accuracy is analyzed. First, the eigenvalues for regions whose closed form solution is known will be compared with the exact solution to demonstrate accuracy. Next, the eigenvalues for which there are only numerical solutions will be presented and their accuracy will be analyzed based on a standard convergence and mesh-refinement analysis. Finally, the most promising candidates for quantum scarring will be presented.

A billiard is a geometric region that is constructed such that the boundaries are perfectly rigid and collisions with it are perfectly elastic. In standard quantum mechanics, the well-known ``particle in a square box" problem is perfectly analogous to a quantum billiard. Although some billiard regions are simple to solve analytically, other more irregular shapes do not have analytical or ``closed form" solutions. As such, numerical methods are integral to finding the eigenvalues of irregularly shaped billiard systems.

We present results for the following billiard shapes: the circle, the equilateral triangle, the stadium billiard, and the 5-pointed star. The circle and equilateral triangle have well-known exact solutions, and as such, they were used as one form of verification for the accuracy of our FEM implementation. All results were obtained using packages available in Wolfram Mathematica; a detailed discussion of these packages will be presented in a later section. The governing equation for the quantum billiards considered here is the helmholz equation: 
\begin{equation}
(\nabla^2 + k^2)\psi = 0,
\end{equation}
which is subject to the boundary conditions imposed by the billiard geometry. A rigorous discussion of the boundary conditions will be presented in the next section.

The eigenfunctions found in conjunction with the eigenvalues were used to generate contour plots within Wolfram Mathematica. For certain state number $n$, quantum scarring is known to emerge in systems that are classically chaotic, such as the stadium billiard. The contour plots generated for this research are examined for indications of quantum scarring. In this paper, we will present a few quantum scarring candidates selected from the numerous contour plots generated. The importance of quantum scarring will be reviewed and further suggestions for improving quantum scarring detection will be discussed.

\section{Numerical Methods}
Perhaps one of the most well-studied problems in quantum mechanics is that of a particle confined to an infinite potential well. The shape of this well, while entirely relevant to the calculation of the allowed energy levels and their corresponding stationary state wave functions, is irrelevant in regard to the overall conceptual understanding of such a region. The regions studied in infinite potential well problems are defined with the following generalized potential:
\begin{equation}
V(r) = \begin{cases}
0, \text{inside the region} \\
\infty, \text{outside of the region}
\end{cases}
\end{equation} \\
Potential regions of this type have \textit{Dirichlet boundary conditions} (hard walls). In terms of the wave function, while inside the region it is nonzero and while outside of the region it cannot exist and is zero-valued. It can be clearly seen that for such a region, a particle is effectively free while inside the boundary. For a free particle, the Schr{\"o}dinger Equation reduces to the Helmholtz equation, which takes the form: \begin{equation}
(\nabla^2 + k^2)\psi(r) = 0
\end{equation}
where $k^2 = 2mE/\hbar^2$~\cite{Kaufman-98}. A system involving a particle in a 2D infinite potential can be termed a \textit{quantum billiard}. For the standard square box and the circular well, analytical solutions are well known and often presented in introductory texts~\cite{Gasiorowicz-03},\cite{Griffiths-18}.
\subsection{The Finite Element Method (FEM)}
The Finite Element Method (FEM) is a numerical method used to solve differential equations over arbitrarily shaped-domains~\cite{Mathematica}. There are four steps to implement FEM~\cite{Wang-16}: \begin{enumerate}
\item Discretize the region and generate a mesh
\item Convert the differential equation of interest to integral form
\item Apply the boundary conditions
\item Assemble system matrices and solve the resulting system of equations
\end{enumerate}
It will be shown that steps 2 and 3 can be carried out simultaneously for Dirichlet boundary conditions. FEM is a natural choice for the modeling of quantum billiards since it is designed to approximate the solution to differential equations over irregularly shaped regions. For clarity, we will briefly review the mathematics behind FEM.

\subsubsection{Region Discretization and Mesh Generation}
We begin implementing FEM by dividing the region of interest into discrete elements in a process called \textit{discretization}. Each discrete element forms the basis of a mesh over the region, and each element need not be the same size~\cite{Wang-16}; the flexibility that results from this form of mesh generation is extremely useful for regions of irregular geometry. FEM seeks to interpolate the solution to a differential equation over a given region. A function $u(x,y)$ that may be the dependent variable in a PDE can be approximated using a linear combination of the basis functions $\varphi_i(x)$, scaled by their nodal values $u_i$:
\begin{equation}
u(x,y) = \sum_{i}u_i\varphi_i(x,y)
\end{equation}
where the nodal value $u_i$ represents the value of the field variable $u(x,y)$ at any given node, and is the expansion coefficient to be determined~\cite{Wang-16}, \cite{Comsol}.

The most basic element in 2D is a triangle, and so the mesh over the domain will have triangular elements, with the vertices as the nodes~\cite{Wang-16}. Therefore, the triangular elements form the surfaces of a 3D object. However, it is important to note that while the simplest elements are triangles, the elements may also take on other shapes, such as rectangles. In simple terms, equation (4) states that the value of the field variable at a particular point in the domain is determined by a basis function scaled by a nodal value that determines its height. For triangular elements, the expansion of $u(x,y)$ consists of three terms that make up the three vertices of the triangular element.

\subsubsection{Converting the Differential Equation to Integral Form}
The next step in FEM is the conversion of the differential equation to an integral form. For a function to be differentiable at a particular point, it must be continuous. However, the boundaries of each element are discontinuous; for the 3D object formed from the linear combination of our basis functions, there is a discontinuity at the transition point between the faces of the object. Therefore, it is necessary to express the differential equation to be solved in its \textit{weak form}~\cite{Wang-16}.

As a specific example, consider the Helmholtz equation:
\begin{equation}
(\nabla^2 + k^2)\Psi(r) = 0 \in \Omega
\end{equation}
where $\Omega$ is the domain in which the differential equation is applied. The application of FEM results in an approximate solution to the differential equation, and as a result there is some associated error with the solution, called the \textit{residual}. To shrink the residual, we weight it by a complex-valued test function $v^*$, and demand that the integral be equal to zero so that the residual vanishes on average in all directions of $v$~\cite{Strang-86}:
\begin{equation}
\int_\Omega v^* (\nabla^2 + k^2)\Psi(r)d\Omega = 0
\end{equation}
Next, by Green's First Identity~\cite{Jackson-62}, we obtain for the laplacian portion of the differential equation: \begin{equation}
\int_\Omega v^*\nabla^2 \Psi(r)d\Omega = \int_{\partial\Omega}v^* \frac{\partial\Psi}{\partial n}d\Gamma - \int_\Omega \nabla v^* \cdot \nabla \Psi d\Omega
\end{equation}
Where $\frac{\partial}{\partial n}$ is the normal derivative at the surface oriented outwards from inside the domain. Combining this result with the $k^2$ term gives the result:
\begin{multline}
\int_\Omega v^*(\nabla^2 + k^2)\Psi(r)d\Omega = -\int_\Omega \nabla v^* \cdot \nabla \Psi d\Omega\\+\int_\Omega v^*k^2\Psi(r)d\Omega~+\int_{\partial \Omega}v^* \frac{\partial \Psi}{\partial n}d\Gamma
\end{multline}

Finally, since the wave function does not exist outside of the domain, the application of boundary conditions yields:
\begin{equation}
\int_\Omega \nabla v^* \cdot \nabla \Psi d\Omega = \int_\Omega v^* k^2 \Psi(r) d\Omega
\end{equation}
Equation (9) is known as the weak form of the differential equation.

\subsubsection{Applying Galerkin's Method}
By this point, equation (9) is still continuous and infinite dimensional. For FEM to work, we need it to be in a discrete form. Therefore, we will apply \textit{Galerkin's Method} to approximate (9). Galerkin's Method has three steps~\cite{Strang-86}: \begin{enumerate}
\item Choose a finite set of trial functions $\phi_i(x,y)$ where $i = 1, ..., n$.
\item Admit approximations to $\psi$ of the form $\displaystyle \Psi(x,y) = \sum_{j=1}^n \psi_j\phi_j(x,y)$
\item Determine the $n$ unknown numbers $\psi_j$ from equation (4.6) using $n$ different test functions $v$ to obtain $n$ equations.
\end{enumerate}
We will choose the test functions to be the same as the trial functions so that $v = \phi_i$ where $i = 1,...,n$. Clearly, the elements constructed in section 1 given by equation (4) satisfy these conditions.

To proceed, we will substitute these approximations into equation (9): \begin{equation}
\sum_{j=1}^{n} \psi_j \left[ \int_\Omega \nabla v^* \cdot \nabla \phi_j(r)d\Omega = \int_\Omega v^* k^2 \phi_j(r)d\Omega \right]
\end{equation}
Recognizing that $v = \phi_j$ and substituting: \begin{equation}
\sum_{j=1}^{n} \psi_j \left[ \int_\Omega \nabla \phi^*_i \cdot \nabla \phi_j(r)d\Omega = \int_\Omega \phi^*_i k^2 \phi_j(r)d\Omega \right]
\end{equation}
The left hand side of equation (11) forms the components of the kinetic energy operator which is symmetric~\cite{Strang-86}: \begin{equation}
K_{ij} = \int_\Omega \nabla \phi^*_i \cdot \nabla \phi_j(r) d\Omega
\end{equation}
The right hand side of equation (11) forms the components of an overlap matrix between the basis functions: \begin{equation}
M_{ij} = \int_\Omega \phi^*_i\phi_j(r)d\Omega
\end{equation}
Substituting these matrices into (11), we arrive at: \begin{equation}
(K_{ij} - k^2M_{ij})\psi_j = 0
\end{equation}

\subsubsection{Assembling the Element Matrices and Solving Linear Equations}
Equation (14) is the component form of an Eigenvalue problem that takes the generalized form:
\begin{equation}
K\psi = k^2M\psi
\end{equation}
and the solutions to Equation (15) provide the discrete eigenvalues and eigenfunctions of interest. $K$ and $M$ are sparse matrices~\cite{Strang-86}, which form a system of linear equations. Since the boundary conditions were previously taken care of in the derivation of the weak form, it need not concern us further here. The system of equations is then solved to obtain the eigenvalues and eigenfunctions. Many eigenvalue problem solvers solve these systems of equations internally, and it is not necessary to go into depth on the details of such computations here. The function NDEigensystem in Mathematica, for example, internally uses iterative Krylov methods such as the Arnoldi Method~\cite{Trefethen-97}.

\subsection{NDEigensystem}
To implement the Finite Element Method in Mathematica, we make use of \textit{NDEigensystem}. NDEigensystem is a function that numerically computes the eigenvalues and eigenfunctions of a differential equation over a specified domain. As input, NDEigensystem takes the specified differential equation (in this case the Helmholtz equation), the dependent variable(s), the domain over which the differential equation is to be solved, and the number of eigenstates desired. 

Although NDEigensystem is designed to handle the PDE discretization itself, we have chosen to generate the mesh separately and then pass the mesh to the function as a region. This was done for the express purpose of preserving user control over the mesh generation process.

The benefit of using NDEigensystem is found in the compact nature of the function. Since the function handles much of the Finite Element Method internally, the code implementation is exceptionally clean; the entire process for finding the eigenvalues for a region and plotting the eigenfunctions can be accomplished in just a few lines of code. Additionally, since the same differential equation is used to model the wave function over different shaped domains, only the domain definition need be changed to model different regions. The result is a highly flexible code implementation that can be repurposed with ease. This flexibility has allowed us to model a variety of regions, the biggest challenge being the limitations of hardware. 

\subsection{Regions Modeled}
Our research modeled numerous regions using NDEigensystem in Mathematica. The first region modeled was a circular billiard. Because the analytical solution is well known for a circular potential well, this region was used to compare the numerical FEM results with the exact solution to ensure precision. The result for the circular billiard was compared with the exact solution given in Kaufman's 1998 paper on modeling quantum billiards with the expansion method~\cite{Kaufman-98}.

The next region modeled was a simple pentagon. We found this region to be useful for assessing the accuracy of FEM for quantum billiard problems. By increasing the number of sides of this polygon to larger and larger values, it can be seen that the eigenvalues of this region converge on the eigenvalues of a circular quantum billiard.

The rest of the regions were chosen to investigate how quantum behavior manifests in a variety of geometries, rather than to model a particular system. The stadium billiard is a traditional region used to study quantum chaos, and a significant amount of time was spent working on modeling this region. Other regions modeled include a 5-pointed star, a 6-pointed star, a generic triangle, an isosceles triangle, an equilateral triangle, and a ``pac-man" shaped region (a circular disk with a sector removed). The computed eigenstates for all regions except the 5-pointed star and stadium billiard are restricted to the first 16 eigenstates; only the stadium and 5-pointed star billiards were investigated for scarring behavior.
\section{Results and Discussion}
The primary focus of this work was to accurately model quantum billiard systems with the Finite Element Method (FEM). In this section, the eigenvalues and eigenfunctions found for each geometric region are presented, and their accuracy is analyzed. First, the eigenvalues for regions whose closed form solution is known will be compared with the exact solution to demonstrate accuracy. Next, the eigenvalues for which there are only numerical solutions will be presented and their accuracy will be analyzed based on a standard convergence/mesh-refinement analysis. Finally, the most promising candidates for quantum scarring will be presented.

\subsection{Closed-Form Solution Comparison}
The exact solutions for the circle and equilateral triangle billiards are provided in \cite{Kaufman-98}. The results from FEM for the circle and equilateral triangle are shown in Table I. The $k$ values are given in atomic units; we will continue to use atomic units throughout the rest of the data. Since the comparison here is only for the first 16 eigenstates, a smaller mesh size was used (max cell measure of 0.0001). As the number of states computed increases, it becomes necessary to increase the mesh size so that the problem remains computationally tractable.
\setlength{\tabcolsep}{8pt}
\begin{table*}[t]
\centering
\caption{Comparison between exact solutions and FEM solutions.\\}
\begin{tabular}{|c|c|c|c|c|c|c|}
\hline
 & \multicolumn{3}{c|}{\textbf{Circle}}
 & \multicolumn{3}{c|}{\textbf{Equilateral Triangle}} \\
\hline
$n$ &
$k_n^{\mathrm{(Exact)}}$ &
$k_n^{\mathrm{(FEM)}}$ &
$\Delta k_n^{\mathrm{(\%)}}$ &
$k_n^{\mathrm{(Exact)}}$ &
$k_n^{\mathrm{(FEM)}}$ &
$\Delta k_n^{\mathrm{(\%)}}$ \\
\hline
1 & 2.4048 & 2.40918 & 0.182136 &
    7.2551 & 7.2552 & 0.001347 \\
\hline
2 & 3.8317 & 3.83865 & 0.181382 &
    11.0824 & 11.0825 & 0.000896 \\
\hline
3 & 3.8317 & 3.83865 & 0.181382 &
    11.0824 & 11.0825 & 0.000896 \\
\hline
4 & 5.1356 & 5.14493 & 0.181673 &
    14.5103 & 14.5104 & 0.000711 \\
\hline
5 & 5.1356 & 5.14493 & 0.181673 &
    15.1028 & 15.1029 & 0.000715 \\
\hline
6 & 5.5200 & 5.53008 & 0.182609 & 
    15.1028 & 15.1029 & 0.000716 \\
\hline
7 & 6.3801 & 6.39173 & 0.182286 &
    18.2585 & 18.2585 & 0.000215 \\
\hline
8 & 6.3801 & 6.39173 & 0.182286 & 
    18.2585 & 18.2585 & 0.00022 \\
\hline
9 & 7.0155 & 7.0283 & 0.182453 &
    19.1954 & 19.1955 & 0.000426 \\
\hline
10 & 7.0155 & 7.0283 & 0.182453 &
    19.1954 & 19.1955 & 0.000428 \\
\hline
11 & 7.5883 & 7.6021 & 0.181859 &
    21.7655 & 21.7657 & 0.000713 \\
\hline
12 & 7.5883 & 7.6021 & 0.181859 &
    22.1649 & 22.1651 & 0.000733 \\
\hline
13 & 8.4172 & 8.4325 & 0.181771 &
    22.1649 & 22.1651 & 0.000741 \\
\hline
14 & 8.4172 & 8.4325 & 0.181771 &
    23.3221 & 23.3223 & 0.000791 \\
\hline
15 & 8.6537 & 8.66941 & 0.181541 &
    23.3221 & 23.3223 & 0.000799 \\
\hline
16 & 8.7714 & 8.78738 & 0.182183 &
    25.4794 & 25.4796 & 0.000607 \\
\hline
\end{tabular}
\end{table*}

It is clear from Table I that the numerical solutions from FEM are consistent with the exact solutions to a high degree. However, it should be noted that this does not unilaterally ensure that the FEM is accurate for other systems and for the eigenvalues associated with high values of the energy level index. Therefore, additional methods of verifying accuracy will be presented.

\subsection{Convergence Test}
To demonstrate that the FEM solutions do indeed converge and give accurate results, a simple procedure was followed for the regions that do not have known analytical solutions. Let $k_h$ be the eigenvalues for a max cell measure $h$. The mesh refinement error is given by: 
\begin{equation}
\epsilon = \frac{|k_h - k_{h/2}|}{k_{h/2}}
\end{equation}
where $k_{h/2}$ is a max cell measure exactly half that of $k_h$. The mesh refinement error is shown for the first 16 eigenvalues in Table II for the stadium billiard and 5-pointed star billiard.

\begin{table*}[t]
\centering
\caption{Mesh Refinement Error for first 16 eigenstates of Stadium and 5-Pointed Star.}
\begin{tabular}{|c|c|c|c|c|c|c|}
\hline
 & \multicolumn{3}{c|}{\textbf{Stadium}}
 & \multicolumn{3}{c|}{\textbf{5-Pointed Star}} \\
\hline
$n$ &
$k_h$ &
$k_{h/2}$ &
$\epsilon$ &
$k_h$ &
$k_{h/2}$ &
$\epsilon$ \\
\hline
1 & 1.78700 & 1.78701 & 3.84370E-07 & 2.06102 & 2.06110 & 3.77452E-05 \\
\hline
2 & 2.30991 & 2.30991 & 8.07064E-07 & 3.22509 & 3.22519 & 3.19709E-05 \\
\hline
3 & 2.96715 & 2.96715 & 9.50551E-07 & 3.22512 & 3.22520 & 2.71827E-05 \\
\hline
4 & 3.27672 & 3.27672 & 3.39308E-07 & 4.10935 & 4.10939 & 9.66233E-06 \\
\hline
5 & 3.64448 & 3.64448 & 8.13079E-07 & 4.10935 & 4.10940 & 1.10937E-05 \\
\hline
6 & 3.67330 & 3.67331 & 1.02531E-06 & 4.44470 & 4.44476 & 1.28346E-05 \\
\hline
7 & 4.16538 & 4.16538 & 1.14838E-06 & 5.17547 & 5.17547 & 8.47751E-07 \\
\hline
8 & 4.39374 & 4.39375 & 1.19971E-06 & 5.17547 & 5.17547 & 8.35119E-07 \\
\hline
9 & 4.76794 & 4.76795 & 1.35504E-06 & 5.32756 & 5.32771 & 2.81324E-05 \\
\hline
10 & 4.81778 & 4.81778 & 7.19331E-07 & 5.32758 & 5.32776 & 3.40705E-05 \\
\hline
11 & 5.05063 & 5.05064 & 1.38890E-06 & 5.96256 & 5.96262 & 9.59901E-06 \\
\hline
12 & 5.18363 & 5.18364 & 1.34652E-06 & 6.28958 & 6.28963 & 8.61723E-06 \\
\hline
13 & 5.40711 & 5.40712 & 1.63477E-06 & 6.28959 & 6.28965 & 1.01990E-05 \\
\hline
14 & 5.52272 & 5.52273 & 1.77803E-06 & 6.61392 & 6.61417 & 3.72050E-05 \\
\hline
15 & 5.88451 & 5.88452 & 1.97582E-06 & 6.61398 & 6.61419 & 3.05980E-05 \\
\hline
16 & 6.05782 & 6.05783 & 2.13442E-06 & 6.92034 & 6.92036 & 3.16541E-06 \\
\hline

\end{tabular}
\end{table*}

Since we are also interested in high energy states, Table III shows the mesh refinement error for each multiple of $n = 100$ so that the mesh refinement error can be examined for large $n$.
\begin{table*}[t]
\centering
\caption{Mesh Refinement Error for multiples of $n = 100$ eigenstates of Stadium and 5-Pointed Star.}
\begin{tabular}{|c|c|c|c|c|c|c|}
\hline
 & \multicolumn{3}{c|}{\textbf{Stadium}}
 & \multicolumn{3}{c|}{\textbf{5-Pointed Star}} \\
\hline
$n$ &
$k_h$ &
$k_{h/2}$ &
$\epsilon$ &
$k_h$ &
$k_{h/2}$ &
$\epsilon$ \\
\hline
100 & 13.99525 & 13.99577 & 3.76871E-05 & 15.75469 & 15.75566 & 6.15568E-05 \\
\hline
200 & 19.52093 & 19.52354 & 1.33662E-04 & 21.75664 & 21.76122 & 2.10412E-04 \\
\hline
300 & 23.76870 & 23.77557 & 2.88779E-04 & 26.39981 & 26.41160 & 4.46403E-04 \\
\hline
400 & 27.23382 & 27.24743 & 4.99437E-04 & 30.34536 & 30.36854 & 7.63660E-04 \\
\hline
500 & 30.37986 & 30.40367 & 7.83971E-04 & 33.82368 & 33.86262 & 1.15132E-03 \\
\hline

\end{tabular}
\end{table*}

Tables II and III show that the mesh refinement error for the lowest energy states is on the order of $10^{-7} - 10^{-5}$, while the highest energy states are on the order of $10^{-3}$. Therefore, it is clear that the eigenvalues have converged and the data is of excellent accuracy. As an additional test of convergence, the eigenvalues of a polygon were examined in the limit where the number of sides increased without bound, effectively approximating a circle billiard. It was found that in this test, the eigenvalues approached that of the circle billiard, reinforcing the accuracy of FEM for billiard systems.

Figure 1 shows the mesh refinement error value as the energy index $n$ increases. As expected, the error increases as the energy index increases. However, even at $n = 500$, the error is within acceptable margins for convergence.

\begin{figure}[!htbp]
  \centering
  \includegraphics[width=1.0\linewidth]{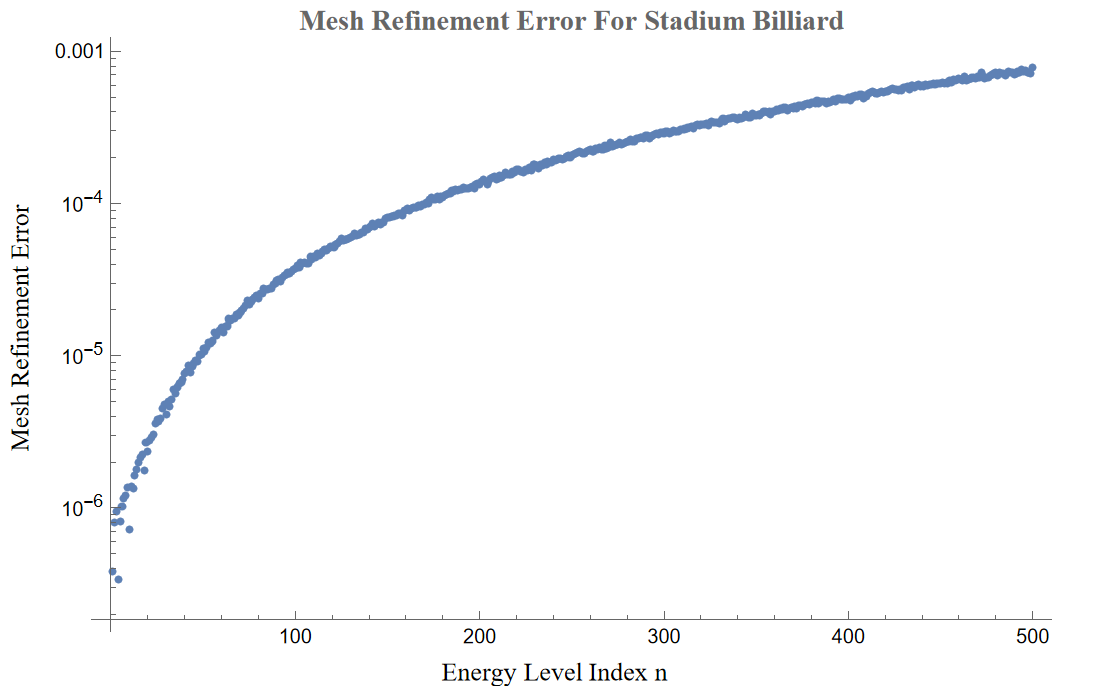}
  \centering
  \includegraphics[width=1.0\linewidth]{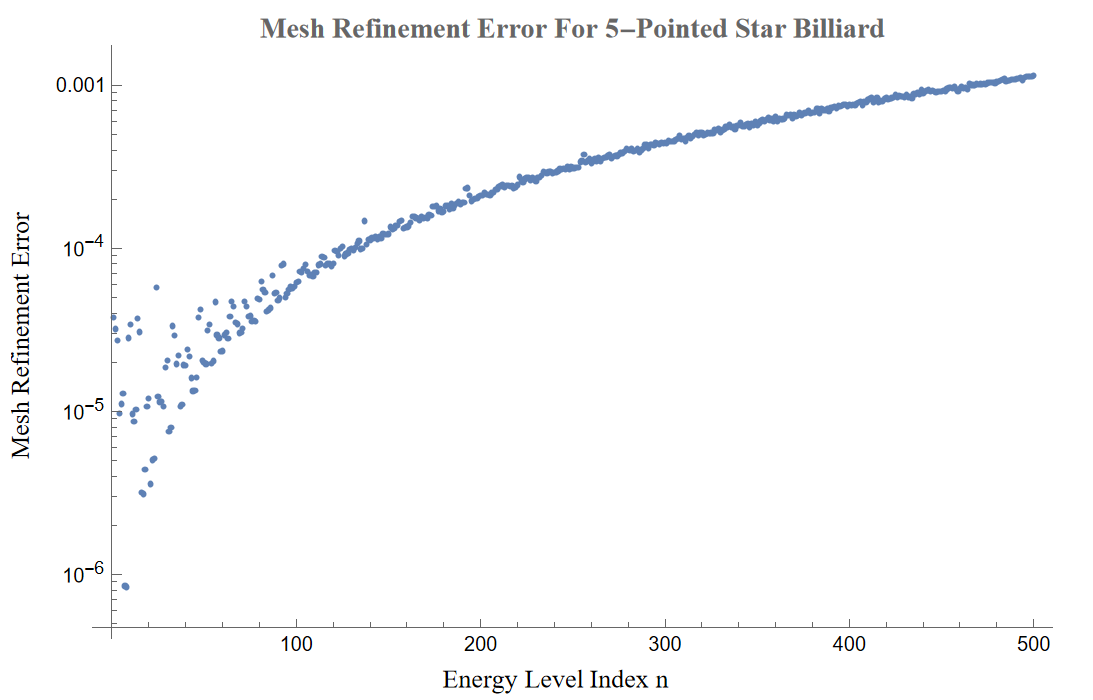}
\caption[Mesh Refinement Error plotted against energy state index $n$ for the Stadium and 5-Pointed Star billiards (semi-log scale).]{Mesh Refinement Error plotted against energy state index $n$ for the Stadium and 5-Pointed Star billiards (semi-log scale).}
\end{figure}

\subsection{Quantum Scarring}
Quantum Scarring is a phenomenon in quantum chaos (quantum dynamics of a classically chaotic system) in which certain eigenstates of a classically chaotic system exhibit unexpected spatial localization along the paths of classical periodic orbits. The conventional expectation is that eigenstates in chaotic systems should resemble random waves and have uniform probability density across the region. However, in 1984 Eric Heller, a researcher at Los Alamos National Laboratory, discovered that a small subset of states show enhanced intensity of the wave function along specific classical trajectories, showing that classical dynamics are encoded in quantum dynamics in a subtle way~\cite{Heller-84}. Since Heller's discovery, research has been done to explore this phenomenon. In this section, we will review the theoretical foundations of quantum scarring, as well as an extension to Heller's quantum scarring definition as made by other researchers~\cite{Bogomolny-04}.

\subsubsection{Heller Scarring}
In a classically chaotic system, the boundary of the system dictates its dynamics which can be integrable, mixed, or fully chaotic~\cite{Das-25}. Among the most famous billiard systems is the well-studied Bunimovich Stadium, which Bunimovich proved to be ergodic, mixing, and Bernoulli, with almost all trajectories exploring the entire accessible region~\cite{Bunimovich-79}; it also has been shown to have unstable periodic orbits (UPOs)~\cite{Biham-92} and Marginally Unstable periodic Orbits (MUPOs)~\cite{Altmann-07}.

Before Heller, it was believed that a quantum chaotic billiard should exhibit a random distribution of the wave function throughout the region in accordance with the Berry Random Wave Model~\cite{Berry-77}. More to the point, the Berry Random Wave Model predicted that high energy eigenfunctions should be statistically featureless, with no visible localization or repeating patterns. However, Heller's research showed a remarkable phenomenon. Rather than a random distribution of the wave function, it instead appeared that certain eigenfunctions were localized along the path of UPOs~\cite{Heller-84}.
\begin{figure}[htbp]
    \begin{center}
        \includegraphics[width = 0.45\textwidth, keepaspectratio]
        {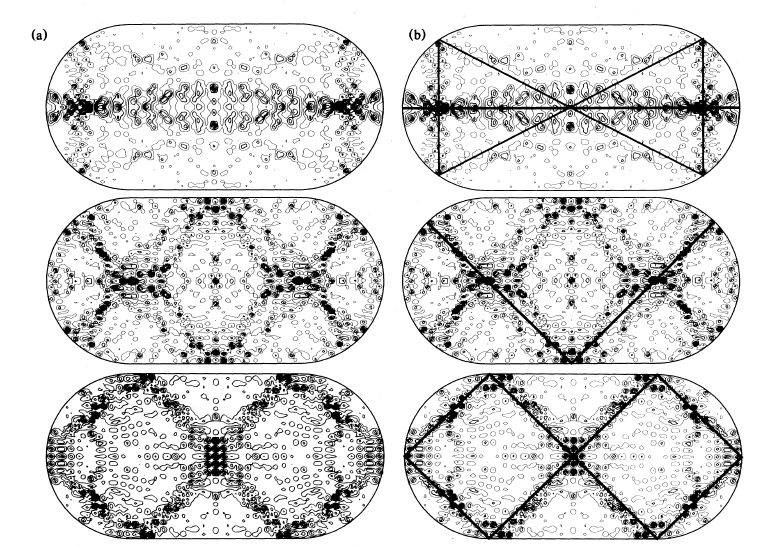}
        \caption{Localization of the wave function along classically UPOs as shown by Heller~\cite{Heller-84}.}
    \end{center}
\end{figure}

Quantum Scarring occurs when a wave packet is launched along a classically unstable periodic orbit. Although the orbit is classically unstable, the wave packet temporarily returns near the orbit periodically. During each recurrence, the wave packet remains partially coherent, so that it constructively interferes with itself. The result is a localized ridge of probability amplitude, which has been termed a quantum scar~\cite{Heller-84}.

\subsubsection{Superscarring}
Heller's definition of quantum scars included only those localizations that occurred along unstable periodic orbits. However, he did recognize that there is another pattern of localization that does not fit into his definition. Heller observed the existence of \textit{non-isolated periodic orbits} bouncing perpendicular to the flat walls of the stadium, which he called a \textit{superscar}~\cite{Heller-84}.

A superscar does not appear to have a strict definition. However, there is a general agreement that superscars have clear structure connected to families of classical periodic orbits which do not disappear at large energies~\cite{Bogomolny-04}. These scars are more persistent than their UPO counterparts. An example of these scars are the bouncing ball modes, one of which can be observed in the stadium billiard bouncing vertically~\cite{Heller-84}.

Although these scars are not necessarily indicative of quantum chaos, they remain an interesting object of study. Superscars, along with traditional quantum scars, are interesting evidence that the behavior of classical systems is subtly encoded into the dynamics of quantum systems. Although there appear to be mathematically rigorous methods for demonstrating the existence of quantum scarring in a system, we will primarily identify \textit{candidates} for quantum scarring via inspection of wave function contour plots.

\subsection{Contour Plots of the Eigenfunctions}
To visualize the eigenfunctions over each region, contour plots were generated. The generation of the contour plots is extraordinarily computationally expensive. This presented a significant challenge in data collection and investigating the region for scars. Additionally, scars are relatively rare which compounded the issue. In the next section, future avenues of research will be discussed along with potential methods of overcoming the computational challenges associated with eigenvalue problems of this nature. Figure 3 shows two examples of typical contour plots generated for the 5-Pointed Star billiard. Although the plots are not indicative of quantum scarring, they exhibit expected symmetries indicating acceptable numerical convergence and symmetry preservation.

\begin{figure}[h!]
\centering
\includegraphics[width=0.8\linewidth]{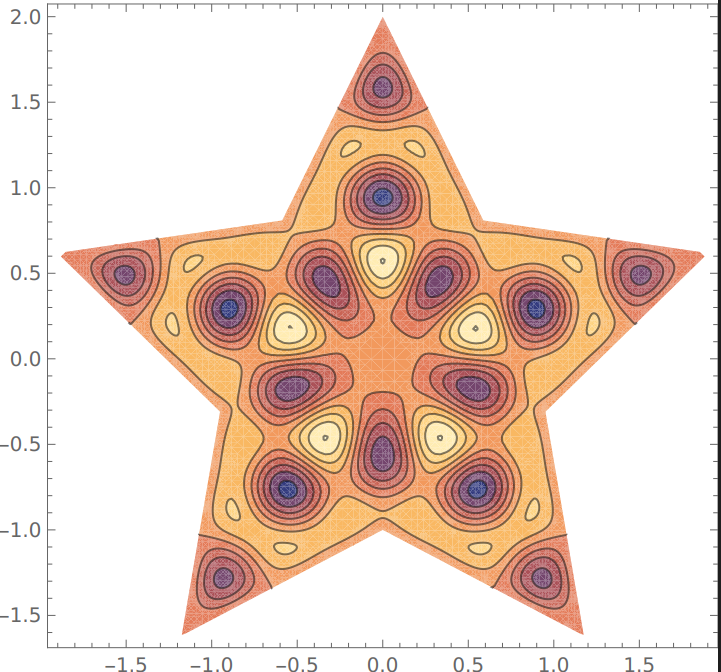}
\centering
\includegraphics[width=0.8\linewidth]{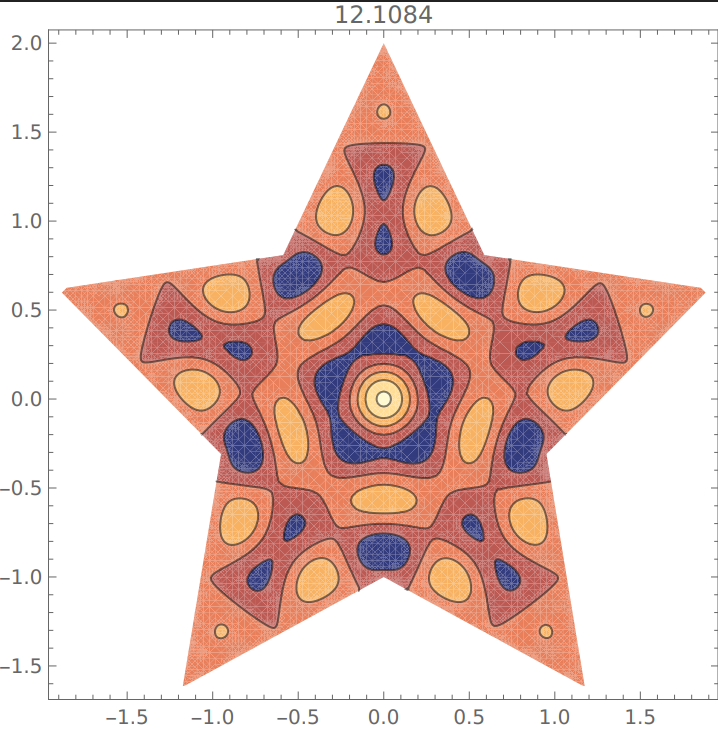}
\caption[Contour Plots for the 5-Pointed Star.]{Contour Plots for the 5-Pointed Star billiard. The top plot corresponds to energy index $n = 49$, and the bottom plot corresponds to energy index $n = 56$.}
\end{figure}

Nevertheless, we have identified a few candidates for quantum scarring and generated numerous plots to compare them against to see the difference between a scar and a typical eigenfunction. The scarring candidates were identified via inspection; there is no standardized method for mathematically identifying quantum scars.

Figure 4 shows two of the most promising scarring candidates found in this research. Although not the traditional scarring Heller investigated, these scarring candidates correspond to vertical bouncing ball modes.

\begin{figure}[h!]
\centering
\includegraphics[width=1.0\linewidth]{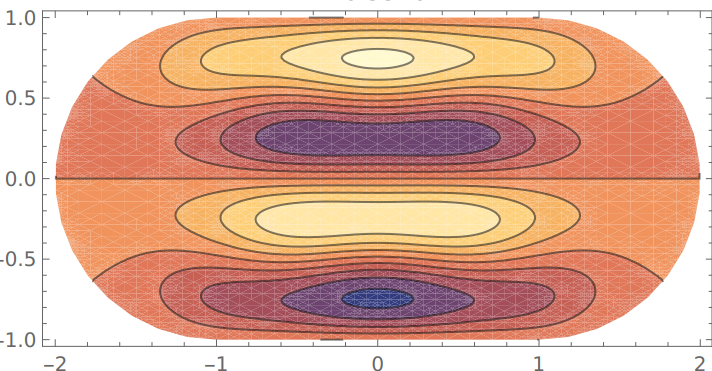}
\centering
\includegraphics[width=1.0\linewidth]{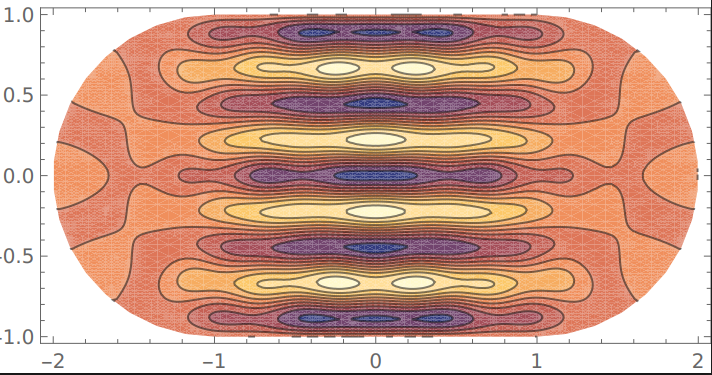}
\caption[Contour Plots for the Stadium Billiard.]{Contour Plots for the Stadium Billiard. The top plot corresponds to energy index $n = 18$, and the bottom plot corresponds to energy index $n = 102$.}
\end{figure}

In the contour plots shown in Figure 4, the wave function appears to be localized along a vertical trajectory in the middle of the stadium. The plot on the left ($n = 18$) is a weak scarring candidate, with the wave function showing some localization but still spread out. The plot on the right however, is a stronger scarring candidate, with the wave function more tightly localized. Despite analyzing over 100 contour plots for the 5-Pointed Star billiard, none of them were considered promising candidates for scarring.

In addition to the scarring candidates presented in figure 4, additional scarring candidates are presented in figure 5. The top left contour plot for the stadium billiard ($n=217$) exhibits wave function localization along the horizontal axis, representative of a horizontal bouncing ball mode. The top right plot ($n=229$) exhibits wave function localization in a "bow-tie" pattern. The bottom left plot ($n = 298$) and the bottom right plot ($n = 293$) are two more types of horizontal bouncing ball scars.

\begin{figure*}[t]
    \centering

    \begin{minipage}[t]{0.48\textwidth}
        \centering
        \includegraphics[width=\linewidth]{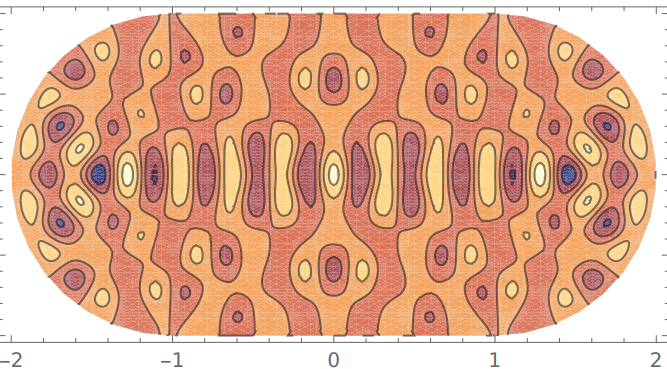}
    \end{minipage}
    \hfill
    \begin{minipage}[t]{0.48\textwidth}
        \centering
        \includegraphics[width=\linewidth]{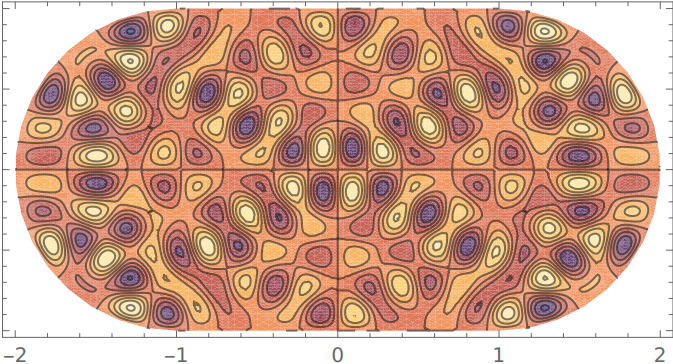}
    \end{minipage}

    \vspace{1em}

    \begin{minipage}[t]{0.48\textwidth}
        \centering
        \includegraphics[width=\linewidth]{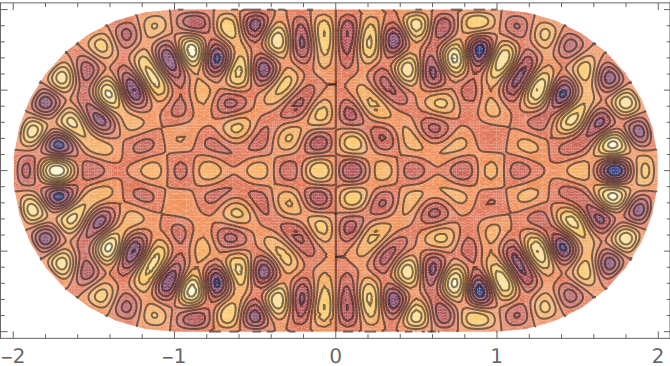}
    \end{minipage}
    \hfill
    \begin{minipage}[t]{0.48\textwidth}
        \centering
        \includegraphics[width=\linewidth]{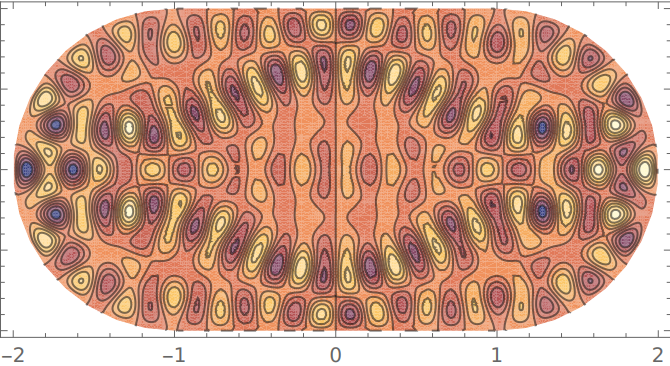}
    \end{minipage}

    \caption[Contour Plots for the stadium billiard.]{Contour plots for the stadium billiard. Top left: $n=217$. Top right: $n=229$. Bottom left: $n=298$. Bottom right: $n=293$.}
\end{figure*}

The scarcity of the quantum scarring phenomenon in the contour plots is not a surprising result. Quantum scars are incredibly rare, and this is reflected in the fact that after reviewing 250 stadium billiard plots and 100 5-Pointed Star billiard plots, only the quantum scarring candidates presented in figures 4 and 5 were identified.

The scarcity of quantum scarring is only part of the challenge. Eigenvalue problems are notoriously difficult for classical computers to solve due to exponential scaling with the size of the system~\cite{Mieldzioc-25}. The enormous amount of time required to compute the eigenvalues and eigenfunctions of a quantum system is not merely a consequence of poor hardware. In fact, even the best classical computers still struggle to evaluate this class of problems; this is a problem that may only truly be resolved with quantum computers and quantum algorithms.

Quantum algorithms such as Quantum Phase Estimation (QPE) are used to find eigenvalues of a given Hamiltonian. Since matrices in quantum eigenvalue problems grow exponentially in size, QPE could in principle help offset the computational cost of solving quantum billiard problems classically.

Additional research can be done to build on the results reported in this paper. A more mathematically rigorous investigation of quantum scarring may be conducted. Additionally, our methods could be improved by employing Adaptive Mesh Refinement (AMR). AMR can increase computational efficiency by automatically adjusting the underlying mesh size to be refined in critical areas and coarsened where the solution is smooth. This research could benefit from AMR as it is computationally expensive to use the same mesh size for a large number of eigenstates. By using AMR, the process of finding the eigenvalues and eigenfunctions for quantum billiards could be optimized. It may also be interesting to investigate whether artificial intelligence could further optimize AMR by recognizing patterns for high-valued eigenstates and identify regions requiring local refinement. Furthermore, the code used for this research is versatile and can be applied to numerous types of geometries. Further research can be done to investigate even more geometric regions, such as the classically chaotic mushroom billiard.

\section{Conclusion}
Since the introduction of quantum theory in the early 20th century, it has revolutionized modern technology. Quantum dots have the potential to be implemented as extremely sensitive sensors, qubits, and as the building blocks for light-based communication. Integral to understanding quantum dots is an understanding of fundamental quantum theory behind its properties, and so quantum billiards have become a useful idealized model to work with for this purpose. 

The FEM is highly effective for evaluating eigenvalues and eigenfunctions for quantum billiard systems. The irregular geometric regions often encountered when studying billiard systems makes FEM a particularly effective method for approximating the eigenvalues and eigenfunctions that result from the Helmholtz equation. 

In addition to the technology that can be created from a deeper understanding of quantum dots through quantum billiards is a deeper understanding of nature itself. The quantum scarring phenomenon present in many billiard systems offers an insight into how classical dynamics emerges from quantum systems, linking two paradigms of physics that all too often seem foreign to one another.

\begin{acknowledgments}
We wish to acknowledge the Massachusetts Space Grant Consortium (MASGC) for the NASA Space Grant Fellowship under which much of the research contained herein was conducted. We also acknowledge Dr. Jianyi Jay Wang, of the University of Massachusetts Dartmouth, for numerous conversations on the subject of quantum scarring which proved invaluable during the data collection process.
\end{acknowledgments}

\nocite{*}

\FloatBarrier
\bibliography{References}

\end{document}